\documentclass[a4paper]{jpconf}
\usepackage{amsmath}
\usepackage{amsfonts}
\usepackage{amssymb}
\usepackage{xcolor}
\usepackage{color}
\usepackage{graphicx}
\begin{document}
\title{Emission of Cosmic Radio--waves, $X$-- or $\gamma$--rays \\
by Moving Unstable Particles at Late Times\footnote{Talk given at \textbf{Seventh International Workshop DICE 2014:}
\textit{\textbf{ Spacetime -- Matter -- Quantum Mechanics
... news on missing links}}, 
Castiglioncello (Tuscany, Italy), September 15 -- 19, 2014 and submitted to the Proceedings of this conference.} }

\author{Krzysztof Urbanowski}

\address{University of Zielona G\'{o}ra, Institute of Physics,
ul. Prof. Z. Szafrana 4a, 65--516 Zielona G\'{o}ra, Poland.}

\ead{K.Urbanowski@if.uz.zgora.pl}

\begin{abstract}
A new quantum effect connected with the late time behavior of decaying states is
described and its possible observational consequences are analyzed:
It is  shown that  charged unstable particles
as well as  neutral unstable
particles with non--zero magnetic moment which
live sufficiently long  may emit electromagnetic
radiation. This mechanism is
due to the nonclassical behavior of unstable
particles at late times (at the
post exponential time region). Analyzing the
transition times region between exponential
and non-expo\-nen\-tial form of the survival
amplitude it  is found that the instantaneous
energy of the unstable particle can take very
large values, much larger than the energy of
this state at times from the exponential
time region. Based on the results obtained
for the  model considered, it is shown that this
new purely quantum  mechanical effect may  be
responsible for causing unstable particles
produced by astrophysical sources and moving
with relativistic velocities
to emit electromagnetic--, $X$-- or
$\gamma$--rays  at some time intervals from  the
transition time regions.
\end{abstract}

\section{Introduction}
Late time properties of unstable states arouse interest since the theoretical
predictions of deviations from the exponential form of the decay law  at suitably
long times. The problem was that the predicted effect is almost negligible smal
and therefore very difficult to its confirmations in laboratory experiments. On the other hand,
numbers of created unstable particles during some astrophysical processes
are so large that
some  of them can survive up to times $t$ at which the survival probability depending on $t$
transforms from  the exponential form into the inverse power--like form.
It appears that at this time region a new quantum effect is observed: A very rapid fluctuations of
the  instantaneous energy of unstable  particles  take place.    These fluctuations of
the instantaneous energy should manifest itself as fluctuations of the velocity
of the particle. Based on results presented in \cite{ku-plb-2014} we show in this report that
this effect may cause unstable particles to emit
electromagnetic radiation of a very wide  spectrum: from radio frequencies through $X$--rays
to $\gamma$--rays of very high energies
and that the astrophysical processes are the place where this effect should be observed.

\section{Preliminaries}

Analyzing properties of unstable states $|\phi \rangle \in {\cal H}$ (where ${\cal H}$ is
 the Hilbert space of states of the considered system)
one usually  starts from studies of their decay law.
The decay law, ${\cal P}_{\phi}(t)$ of
an unstable state $|\phi\rangle$ decaying
in vacuum
is defined as follows
\begin{equation}
{\cal P}_{\phi}(t) = |a(t)|^{2}, \label{P(t)}
\end{equation}
where $a(t)$ is  the probability amplitude of finding the system at the
time $t$ in the initial state $|\phi\rangle$ prepared at time $t_{0}
= 0$,
\begin{equation}
a(t) = \langle \phi|\phi (t) \rangle . \label{a(t)}
\end{equation}
and $|\phi (t)\rangle$ is the solution of the Schr\"{o}dinger equation
for the initial condition  $|\phi (0) \rangle = |\phi\rangle$:
 \begin{equation}
i\hbar \frac{\partial}{\partial t} |\phi (t) \rangle = H |\phi (t)\rangle,  \label{Schrod}
\end{equation}
 where $H$ denotes the total self--adjoint Hamiltonian for the system considered.

 From basic principles of quantum theory it is known that the
amplitude $a(t)$, and thus the decay law ${\cal P}_{\phi}(t)$ of the
unstable state $|\phi\rangle$, are completely determined by the
density of the energy distribution $\omega({\cal E})$ for the system
in this state \cite{fock}
\begin{equation}
a(t) = \int_{Spec.(H)} \omega({\cal E})\;
e^{\textstyle{-\frac{i}{\hbar}\,{\cal E}\,t}}\,d{\cal E}.
\label{a-spec}
\end{equation}
where $\omega({\cal E}) \geq 0$  and $a(0) = 1$.

In \cite{khalfin}
assuming that the spectrum of $H$ must be bounded
from below, $(Spec.(H) = [E_{min},\infty)$ and $E_{min} > -\infty)$, and using the Paley--Wiener
Theorem  \cite{paley}
it was proved that in the case of unstable
states there must be
$|a(t)| \; \geq \; A\,\exp\,[- b \,t^{q}]$,
for $|t| \rightarrow \infty$. Here $A > 0,\,b> 0$ and $ 0 < q < 1$.
This means that the decay law ${\cal P}_{\phi}(t)$ of unstable
states decaying in the vacuum, (\ref{P(t)}), can not be described by
an exponential function of time $t$ if time $t$ is suitably long, $t
\rightarrow \infty$, and that for these lengths of time ${\cal
P}_{\phi}(t)$ tends to zero as $t \rightarrow \infty$  more slowly
than any exponential function of $t$.
This effect was confirmed
in experiment described  in the Rothe paper \cite{rothe},
where the experimental evidence of deviations from the exponential decay law at long times was
reported. In the light of this result the following problem seems to be worth considering:
If (and how) deviations from the
exponential decay law at long times affect the energy of the unstable state
and its decay rate at this time region.

\begin{figure}[h!]
\begin{center}
\includegraphics[width=120mm]{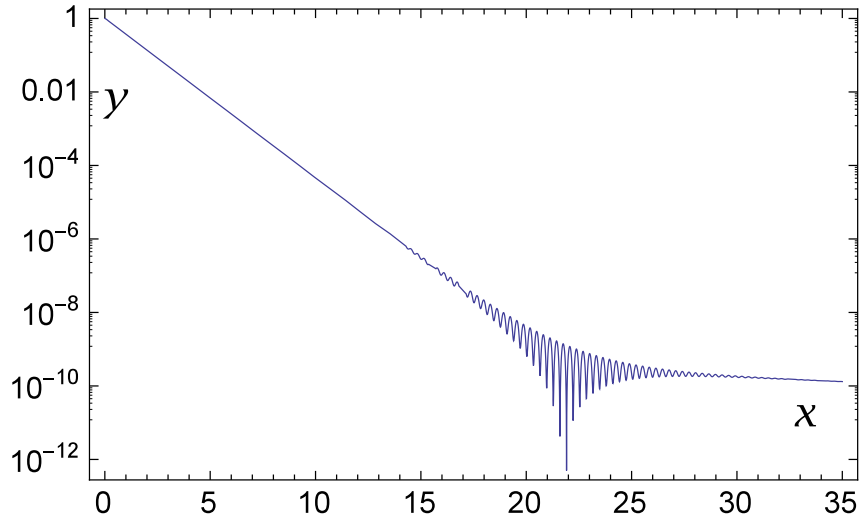}
\caption{The general, typical  form of the decay curve ${\cal P}_{\phi}(t)$.
Axes: $x =t / \tau_{\phi}, y = {\cal P}_{\phi}(t)$ --- the logarithmic scale. }
\label{fa}
\end{center}
\end{figure}

Studying properties of unstable states at late times
it is useful to express  $a(t)$  in the
following form
$a(t) = a_{exp}(t) + a_{lt}(t)$,
where $a_{exp}(t)$ is the exponential part of $a(t)$, that is
$a_{exp}(t) =
N\,\exp\,[{-i\frac{t}{\hbar}(E_{\phi}^{0} - \frac{i}{2}\,{\it\Gamma}_{\phi}^{0})}]$,
($E_{\phi}^{0}$ is the energy of the system in the state $|\phi\rangle$
measured at the canonical decay times,
i.e. when ${\cal P}_{\phi}(t)$ has the exponential form,
${\it\Gamma}_{\phi}^{0}$ is the decay width, $N$ is the normalization
constant), and $a_{lt}(t)$ is the
late time non--exponential part of $a(t)$.
From the literature it is known  that the characteristic  feature of
survival probabilities ${\cal P}_{\phi}(t)$ is the presence of sharp and frequent fluctuations at
the transition times region, when contributions from $|a_{exp}(t)|^{\,2}$ and
$|a_{lt}(t)|^{\,2}$ into ${\cal P}_{\phi}(t)$ are comparable
and that the amplitude  $a_{lt}(t)$ and thus the probability ${\cal P}_{\phi}(t)$ exhibits inverse
power--law behavior at the late time region for times $t$ much later than the
crossover time $T$.
The crossover time $T$
can be found by solving the following equation,
\begin{equation}
|a_{exp}(t)|^{\,2} = |a_{lt}(t)|^{\,2}. \label{T}
\end{equation}
In general $T \gg \tau_{\phi}$, where
$\tau_{\phi} = \hbar/{\it\Gamma}^{0}_{\phi}$ is the live--time of $\phi$.
Formulae for $T$ depend on the model considered (i.e. on $\omega(E)$) in general).
A typical form of the decay curve, that is the form of the non--decay probability ${\cal P}_{\phi}(t)$
as a function of time $t$ is presented in Fig. (\ref{fa}).
The results of calculations presented in this Figure  were obtained for
the  Breit--Wigner   energy distribution function,
$\omega ({E}) \equiv \omega_{BW}(E)$, where
\begin{equation}
\omega_{BW}(E) \stackrel{\rm def}{=}
 \frac{N}{2\pi}  {\it\Theta} ( E - E_{min})
\frac{{\it\Gamma}_{\phi}^{0}}{( E - E_{\phi}^{0})^{2} +
(\frac{{\it\Gamma}_{\phi}^{0}}{2})^{2}}, \label{omega-BW}
\end{equation}
and $\it\Theta ({E})$ is the unit step function.
The calculations leading to  graphics presented in Fig. (\ref{fa})
were performed for $s_{0} \stackrel{\rm def}{=}({E_{\phi}^{0}} - E_{min}) / {\it\Gamma}_{\phi}^{0}  $ $= 20$.

\section{Late times and energy of unstable states}
The energy and the decay rate of the system in the state $|\phi\rangle$
under considerations, (to be more precise  the instantaneous energy and the instantaneous decay rate),
can be found using the following relations
(for details see \cite{ku-cejp-2009,ku-ijmpa-1991,ku-epjd-2009}).
\begin{equation}
{\cal E}_{\phi} \equiv {\cal E}_{\phi}(t) = \Re\,(h_{\phi}(t)),\;\;\;\;
\gamma_{\phi} \equiv \gamma_{\phi}(t) = -\,2\,\Im\,(h_{\phi}(t)),
\label{E(t)}
\end{equation}
where $\Re\,(z)$ and $\Im\,(z)$ denote the real and imaginary parts
of $z$ respectively and  $h_{\phi}(t)$ denotes the "effective
Hamiltonian" for the one--dimensional subspace of
states ${\cal H}_{||}$ spanned by the normalized vector
$|\phi\rangle$,

\begin{equation}
h_{\phi}(t) \stackrel{\rm def}{=}  i \hbar\, \frac{\partial
a(t)}{\partial t} \; \frac{1}{a(t)}. \label{h}
\end{equation}
It is easy to show that equivalently
\begin{equation}
h_{\phi} (t) \equiv \frac{\langle \phi|H|\phi (t)\rangle}{\langle \phi |\phi (t)\rangle}.
\label{h-equiv}
\end{equation}

There is ${\cal E}_{\phi}(t)= E_{\phi}^{0}$ and ${\it\Gamma}_{\phi}(t) =
{\it\Gamma}_{\phi}^{0}$ at the canonical decay
times   and at asymptotically late
times \cite{ku-plb-2014,ku-cejp-2009,ku-epjd-2009,ku-prl-2011}
\begin{eqnarray}
{\cal E}_{\phi}(t) &\simeq& E_{min} + \frac{c_{2}}{t^{2}}
\,+\,\frac{c_{4}}{t^{4}} \ldots, \;\;\;({\rm for}
\;\;t \gg T), \label{E(t)} \\
\it\Gamma_{\phi}(t)  &\simeq& \frac{c_{1}}{t}
+ \frac{c_{3}}{t^{3}} + \ldots \;\;\;({\rm for}
\;\;t \gg T), \label{g(t)}
\end{eqnarray}
where $c_{i} = c_{i}^{\ast},\;i=1,2,\ldots$, ($c_{1} > 0$ and
the sign of $c_{i}$ for $i \geq 2$ depends on the model considered),
so $\lim_{t \rightarrow \infty}\, {\cal E}_{\phi}(t) = E_{min}$
and $\lim_{t \rightarrow \infty}\, {\it\Gamma}_{\phi}(t) = 0$.
Results (\ref{E(t)}) and (\ref{g(t)}) are rigorous.

The basic physical factor forcing
the  amplitude $a(t)$ to exhibit inverse power law behavior at $t \gg T$ is the
boundedness from below of  $\sigma (H)$. This means that if this condition is satisfied
and $\int _{-\infty}^{+\infty}\omega(E)\,dE\,\equiv \,\int _{E_{min}}^{+\infty}\omega(E)\,dE\,<\,\infty$, then all  properties
of $a(t)$, including the  form of the time--dependence at  times $t \gg T$, are the
mathematical consequence of them both.
The same applies by (\ref{h}) to the properties of $h_{\phi}(t)$ and concerns
the asymptotic form of $h_{\phi}(t)$ and
thus of ${\cal E}_{\phi}(t)$ and $\it\Gamma_{\phi}(t)$ at $t \gg T$.

\begin{figure}[h!]
\begin{center}
\includegraphics[width=120mm]{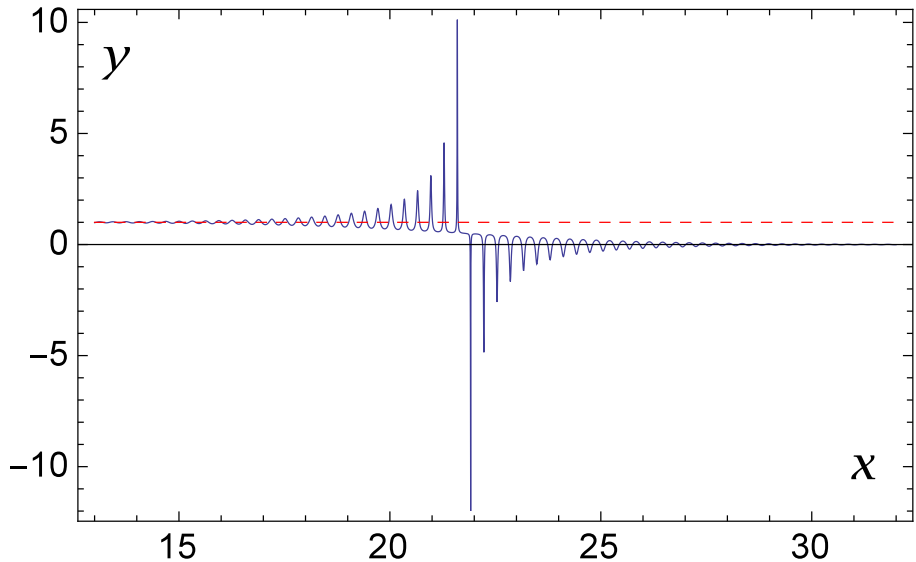}
\caption{The instantaneous energy ${\cal E}_{\phi}(t)$
  in the transitions time region:  The case $ s_{0} = 20$.
 Axes: $y= \kappa(t)$, $x =t / \tau_{\phi}$.
 The dashed line denotes the  straight line $y=1$.}
  \label{f1}
\end{center}
\end{figure}

Using relations (\ref{a(t)}), (\ref{a-spec}), (\ref{h}) and assuming the form of $\omega(E)$ and
performing all necessary calculations numerically one can find ${\cal E}_{\phi}(t)$ for all
times $t$ including times $t \gg T$.
A typical behavior  of the instantaneous energy ${\cal E}_{\phi}(t)$
at the transition time region is presented in Fig (\ref{f1}).
In this Figure
calculations were performed for  the  Breit--Wigner
energy distribution function (\ref{omega-BW}) and function $\kappa (t)$ is defined as follows
\begin{equation}
\kappa (t) = \frac{{\cal E}_{\phi}(t) - E_{min}}{ E_{\phi}^{0} - E_{min}}. \label{kappa}
\end{equation}

As it can be seen in Fig. (\ref{fa})
the characteristic  feature  decay curves is the
presence of sharp and frequent oscillations at the transition times region.
This means that derivatives of the amplitude $a(t)$ may reach extremely large
values for some times from the transition time region and the modulus of these derivatives is much
larger than the modulus of $a(t)$, which is very small for these times. This explains why in this time
region the real and imaginary parts of $h_{\phi}(t) \equiv {\cal E}_{\phi}(t) \, - \,\frac{i}{2}\,\gamma_{\phi}(t)$,
which can be expressed by the relation (\ref{h}), ie. by a large
derivative of $a(t)$  divided by a very small $a(t)$, reach values much larger than the
energy ${\cal E}_{\phi}^{0}$ of the the unstable state measured at times for which the decay
curve has the exponential form.

\section{Possible observable effects}
From the point of view of a frame of reference  in which the time evolution
of the unstable system was   calculated the Rothe experiment as well as the
picture presented in Figs (\ref{fa}), (\ref{f1}) refer to the rest coordinate system of the
unstable system considered.

Properties of the ratio $\kappa (t)$ taking place
for some time intervals mean
that the  instantaneous energy ${{\cal E}_{\phi}(t)}$ at these time intervals differs
from the energy $ E^{0}_{\phi}$ measured at
times from the exponential decay time region (i.e. at the canonical decay times).
The relation (\ref{h-equiv})
explains why such an effect can occur. Indeed, if to rewrite the numerator
of the righthand side of (\ref{h-equiv}) as follows,
\begin{equation}
\langle \phi|H|\phi (t)\rangle \equiv \langle
\phi|H|\phi\rangle\,a(t)\,+\,\langle \phi |H|\phi (t)\rangle_{\perp}, \label{perp}
\end{equation}
where $|\phi (t)\rangle_{\perp} = Q|\phi (t)\rangle$,
$Q = \mathbb{I} - P$ is the projector onto the subspace od decay products,
$P = |\phi\rangle\langle \phi|$ and $\langle \phi|\phi (t)\rangle_{\perp} = 0$,
then one can see
that decay products described by $|\phi (t)\rangle_{\perp}$ contribute
permanently to the energy of the unstable state considered with the intensity depending on time $t$.
This contribution into the instantaneous energy is practically constant
in time at  canonical decay times
whereas at the transition times, when $t \sim T$, it
is fluctuating function of time and the amplitude of these fluctuations may be significant.

From astronomical observations it is known that
there are sources of unstable particles in space that emit
them  with relativistic or ultra--relativistic velocities in relation to an external
observer so many of these particles move in space  with ultra high energies.
In such cases one has to consider the following problem: What effects
can be observed by an external observer when the unstable
particle, say $\phi$,   which survived up to the transition times region, $t\sim T$, or longer
is moving with a relativistic velocity in relation to this observer.
The distance $d$ from the source reached by this particle is of order
$d \sim d_{T}$,  where
$ d_{T} = v^{\phi} \cdot T'$
and $T' = \gamma \, T$, $\gamma \equiv \gamma (v^{\phi})= (\sqrt{1-\beta^{2}})^{-1}$, $ \beta = v^{\phi}/c$,
$v^{\phi}$ is the velocity of the particle  $\phi$. (For simplicity we assume that
there is  a frame of reference common for the source and observer
both and that they do not move with respect to this frame of reference). So,
in the case of moving particles created in astrophysical processes one should consider the
effect shown in Fig (\ref{f1}) together with   the fact that the particle
gains extremely huge  energy, $W^{\phi}$,
which have to be conserved.

Let us assume that the unstable particle under considerations has a constant momentum $\vec{p}$, $|\vec{p}|=p >0$.
Next, let us assume for simplicity that
$\vec{v} = (v_{1},0,0) \equiv (v^{\phi},0,0)$ then there is $\vec{p}=(p,0,0)$.
Now, let $\Lambda_{p}$ be the Lorentz transformation from the reference
frame ${\cal O}$, where the momentum of unstable particle considered is zero,
$\vec{p}=0$, into the frame ${\cal O}'$ where  the momentum of this
particle is $\vec{p} \equiv (p,0,0) \neq 0\;$
or, equivalently, where its velocity $\vec{v}^{\,\phi}$ equals $\vec{v}^{\,\phi} = \vec{v}_{p}^{\phi} \equiv
\vec{p}/ m\gamma$,
and let $E$ be the energy of the particle in the reference frame ${\cal O}$ and $E'$  be the energy of the moving particle
and having the  momentum $\vec{p} \neq 0$. In this case the corresponding
4--vectors are: $\wp =(E/c,0,0,0)  \in {\cal O}$
and   $\wp'=(E'/c,p,0,0)
 = \Lambda_{p}\,\wp \,\in \,{\cal O}'$.
There is $\wp'\cdot \wp' \equiv (\Lambda_{p} \wp)\cdot (\Lambda_{p} \wp)
= \wp\cdot \wp$
in Minkowski space, which is an effect  of the  Lorentz
invariance.
(Here the dot "$\cdot$" denotes the scalar product in Minkowski space).
Hence, in our case:
$\wp'\cdot \wp' \equiv (E'/c)^{2} - p^{2} = (E/c)^{2}$
because
$\wp \cdot \wp \equiv (E/c)^{2}$
and thus
$(E')^{2} = {c^{2}\,p^{2} + E^{2}}$.
One can show that in the case of a free particle $p\equiv  \frac{E}{c}\,\gamma\, \beta $. Therefore
$c^{2}\,p^{2} + E^{2} \equiv E^{2} \,\gamma^{2}\,\beta^{2}\; +\; E^{2} = E^{2}(1\,+\,\beta^{2}\,\gamma^{2}) \equiv E^{2}\,\gamma^{2}$.
The last relation means that  the energy, $E'$, of moving particles equals
\begin{equation}
E' \,=\, \sqrt{E^{2} + c^{2}\,p^{2}} \equiv E\,\gamma \stackrel{\rm def}{=} W^{\phi}.\label{E'2}
\end{equation}
So in  the  case of the instantaneous energy, ${\cal E}_{\phi}(t)$ considered earlier  we can write that
\begin{equation}
W^{\phi}  \equiv {\cal E}_{\phi}(t_{k})\,\gamma_{k}  = W^{\phi}(t_{k}),\label{W-phi}
\end{equation}
at the instant $t_{k}$ (where $\gamma_{k} = \gamma (t_{k})$.

\begin{figure}[h!]
\begin{center}
\includegraphics[width=110mm]{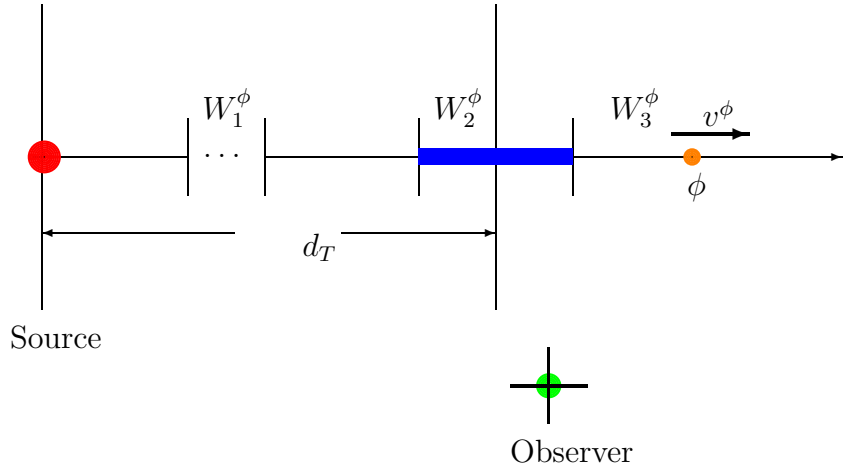}
\caption{Here $d_{T} = v^{\phi} \cdot T',\, T' = \gamma \, T,\; W_{i}^{\phi} = W^{\phi}(t_{i}),\;(i=1,2,3)$ and $W_{i}^{\phi}(t_{i})$
is the energy of moving relativistic particle $\phi$, $t_{1} \ll t_{2} \ll t_{3}$,
$ W^{\phi}(t_{i}) = {\cal E}_{\phi}(t_{i})\,\gamma (t_{i})$,  $\gamma = \gamma (t) = (\sqrt{1-\beta^{2}})^{-1}$, $ \beta = v^{\phi}(t)/c$. }
\label{F3}
\end{center}
\end{figure}

 We have
${\cal E}_{\phi}(t) \equiv E_{\phi}^{0}$ at canonical decay times. Thus,
$W^{\phi} \,\equiv \,E_{\phi}^{0} \, \gamma$,
  at these times  (where $\gamma = \gamma (t \sim \tau_{\phi}$) ) and
at times $t \gg \tau_{\phi}, \; t \sim T$
we have ${\cal E}_{\phi}(t) \neq E_{\phi}^{0}$.
Invoking now the principle of conservation of energy, we can conclude that
in the case illustrated in Fig (\ref{F3}), we have
\begin{equation}
W_{1}^{\phi}= W_{2}^{\phi} = W_{3}^{\phi},
\label{w1=w2}
\end{equation}
where $W_{i}^{\phi} = W^{\phi}(t_{i})$.

The most interesting relation can be obtained  from (\ref{w1=w2})
for $t_{1} \ll T'$ and  $t_{2} \sim T'$, (where $T'= \gamma\,T$): $W_{1}^{\phi}\equiv E^{0}_{\phi}\,\gamma\, =\, W_{2}^{\phi}$,
or the equivalent one,
\begin{equation}
 { E}^{0}_{\phi} \,\gamma_{1} \equiv {\cal E}_{\phi}(t_{2})\,\gamma (t_{2}). \label{W1W2}
\end{equation}
It is convenient to rewrite the last relation as follows,
\begin{equation}
\gamma_{1} =  \frac{{\cal E}_{\phi}(t_{2})}{{E}^{0}_{\phi}} \gamma (t_{2}).
\label{g1=kg2a}
\end{equation}
The ratio $\frac{{\cal E}_{\phi}(t_{2})}{{E}^{0}_{\phi}}$ can be extracted from Fig (\ref{f1}):
\begin{equation}
\frac{{\cal E}_{\phi}(t)}{ E_{\phi}^{0}} =
1\, + \,
\big(\kappa (t)\,-\,1\big)\,\frac{E_{\phi}^{0} - E_{min}}{E_{\phi}^{0}}. \label{E/E0}
\end{equation}
Thus if $ \kappa (t)  \neq 1 $ then $\gamma_{2} \neq \gamma_{1}$, which
means that $v_{2}^{\phi} \neq v_{1}^{\phi}$ or that $a_{average} =
\frac{v_{2}^{\phi} - v_{1}^{\phi}}{t_{2} - t_{1}} \neq 0$.
So,  the moving charged unstable particle $\phi$ has to emit electromagnetic
radiation at the transition time region. The same concerns neutral unstable
particles with non--zero magnetic moment. The energy of the electromagnetic radiation
emitted by such a charged particle in a unit of time can be found using
the Larmor formula: 

\pagebreak

\begin{equation}
P = \frac{1}{6 \pi \epsilon_{0}}\,\frac{q^{2}\, {\dot{v^{\phi}}}^{2}}{c^{2}}\,\gamma^{6},
\label{dwdt}
\end{equation}
where $P$ is total radiation power, $q$ is the electric charge,
$ \epsilon_{0}$ -- permittivity for free space.
The known analogy from classical physics is the following one:
A conservation of the angular momentum and a pirouette like effect.

\section{Some estimations}
From the results presented in
Fig (\ref{f1}) one can find coordinates of, e.g., the highest maximum:
$(x_{mx},y_{mx}) = (21.60,10.27)$.  They allow one to find that $\kappa (t_{mx}) = y_{mx} = 10.27$.
Coordinates of points of the intersection of this maximum with the straight line $y=1$ are equal:
$(x_{1}, y_{1})
= (21.58, 1.0)$ and $(x_{2}, y_{2}) = (21.62, 1.0)$.
Thus
$\Delta x  = x_{2} - x_{1} = 0.04$. (Here $x = t/\tau_{\phi}$),
and this $\Delta x$ allows one to find  $\Delta t =t_{mx} - t_{1}$.
Let us consider now $\mu$ meson as an example.
In such a case: $E_{\mu}^{0} - E_{min} = m_{\mu^{\pm}} - (m_{e}
+ m_{\bar{\nu}_{e}} + m_{\nu_{\mu}}) \simeq 105$ [MeV],
($\tau_{\mu} = 2,198 \times 10^{-6} [{\rm s}]$ and the crossover time equals:
$T^{\mu} \simeq 165 \tau_{\mu} = 0,37 \times 10^{-3} [{\rm s}], \;\;
$ \cite{ku-plb-2014}). Using these data one finds that $\gamma(t_{1}) \simeq 10.21\gamma (t_{mx})$.
From this relation one can extract  $\Delta v^{\mu}$. Next using such
obtained $\Delta v^{\mu}$ and $\Delta t =t_{mx} - t_{1}$ extracted from
Fig (\ref{f1}) one can use the Larmor formula (\ref{dwdt}) and to calculate
the  total radiation power $P$. The result is $P \sim 4.6$ [eV/s].

Astrophysical and cosmological processes in which  extremely
huge numbers of unstable particles are created
seem to be  a possibility for the above discussed
effect to become manifest.
The fact is that
the probability ${\cal P}_{\phi}(t)
= |a(t)|^{2}$ that an  unstable particle $\phi$ survives
up to time $t \sim T$ is extremely small.
According to estimations of the luminosity of some
$\gamma$--rays sources the energy  emitted  by  these sources  can even
reach a value of order $10^{52}$ [erg/s],
and it is only a part of the total energy produced there.
If to consider a source
emitting energy $10^{50}$ [erg/s] then, eg., an emission of
${\cal N}_{0} \simeq 6.25 \times 10^{47}$ [1/s] particles of energy
$10^{18}$ [eV] is energetically allowed. The same source can emit
${\cal N}_{0} \simeq 6.25 \times {10}^{56}$ [1/s] particles of energy $10^{9}$ [eV] and so on.

For  the model defined by
$\omega_{BW}(E)$
the cross--over time $T$ is given by the following approximate
relation (valid for ${E_{\phi}^{0}}/{\it\Gamma_{\phi}^{0}} \gg 1 $),
\begin{equation}
 {\it\Gamma}_{\phi}^{0}\,T \equiv \frac{T}{\tau_{\phi}} \;
 \sim\;2\, \ln \,\Big[2 \pi \Big(\frac{E_{\phi}^{0} - E_{min}}{\it\Gamma_{\phi}^{0}}\Big)^{2}\Big],
\label{t-as}
\end{equation}
(see, eg. \cite{ku-cejp-2009,ku-epjd-2009,ku-epjc-2008}).
As it was mentioned, e.g. in \cite{krauss}, a  typical
value of the ratio ${(E_{\phi}^{0}} - E_{min})/{{\it\Gamma}_{\phi}^{0}}$ is
$({E_{\phi}^{0}}- E_{min})/{{\it\Gamma}_{\phi}^{0}} \,
\geq \, O (10^{3} - 10^{6})$.
Taking
$({E_{\phi}^{0}}-E_{min})/{{\it\Gamma}_{\phi}^{0}} =
10^{6}$ one obtains from (\ref{t-as}) that
${\cal N}_{\phi}(T) \equiv {\cal P}_{\phi}(T) \,{\cal N}_{0} \sim 2.53 \times 10^{-26}\, {\cal N}_{0}$.
%\hfill\\
So there are ${\cal N}_{\phi}(T) \sim 14 \times 10^{21}$ particles per second of
energy $W^{\phi} = 10^{18}$ [eV] or ${\cal N}_{\phi}(T) \sim 14 \times 10^{30}$ particles
of energy $W^{\phi} = 10^{9}$ [eV] in the case of the considered example and $T$
calculated using (\ref{t-as}).

\section{Conclusions}
\begin{itemize}
\item
Fluctuations of the instantaneous energy, ${\cal E}_{\phi}(t)$,
of the moving unstable particles  at transition times region,
$t \sim T$, (where $T$ is the crossover time --- see (\ref{T}) and (\ref{t-as}))
cause fluctuations of the velocity of these particles at this time
region, which forces the charged particles to emit electromagnetic radiation.
\item
Similarly to the charged particle, a moving  neutral unstable particle with non--zero magnetic
moment have to emit electromagnetic radiation in the transition time region.
\item
Ultra relativistic unstable charged particles or unstable particles with non--zero
magnetic moment can emit $X$-- or $\gamma$--rays in the transition time region.
\item
Astrophysical sources are able to create such numbers ${\cal N}_{0}$ of unstable
particles that sufficiently large number ${\cal N}_{\phi}(T) \gg 1$ of them has to survive up to
times $T$ and therefore to emit electromagnetic radiation at transition times.
The expected spectrum of this radiation can be very wide:
From radio frequencies up to $\gamma$--rays.
\end{itemize}

\ack{
The work was supported by the NCN grant No
DEC-2013/09/B/ST2/03455.}

\end{document}